\documentclass[6pt, preprint2]{aastex}
\begin{document}

\title{Evidence in Support of the Local Quasar Model from Inner Jet Structure and Angular Motions in Radio Loud AGN}

\author{M.B. Bell\altaffilmark{1}}

\altaffiltext{1}{Herzberg Institute of Astrophysics,
National Research Council of Canada, 100 Sussex Drive, Ottawa,
ON, Canada K1A 0R6;
morley.bell@nrc.gc.ca}

\begin{abstract}

Radio loud jetted sources with and without extended inner jet structure show good agreement with the simple ballistic ejection scenario proposed in the decreasing intrinsic redshift (DIR) model, where, because of projection effects, those that show the most obvious extended structure and large angular motions are assumed to have jets that lie close to the plane of the sky, and those with little or no structure and small angular motions are assumed to have jets that are coming almost directly towards us. This simple model also predicts several other relations seen in the raw data that, in some cases, may be less easily explained if the redshifts are cosmological and relativistic ejection is required. In particular, for radio-loud sources the source number density is found to be high for sources that are not Doppler boosted but low for highly boosted sources. This is opposite to what is expected, \em suggesting that Doppler boosting may not be involved at all, \em which would be in agreement with the DIR model. If so, the reality of relativistic beaming in quasar jets, the assumption of which has been the very foundation of the superluminal motion explanation in the cosmological redshift (CR) model, would then be questioned.

 

\end{abstract}

\keywords{galaxies: active - galaxies: distances and redshifts - galaxies: quasars: general}

\section{Introduction}

The large apparent superluminal motions observed in the jets of many radio-loud AGN galaxies have been explained in the CR model by assuming that the ejected blobs are moving towards us with near light speeds and with small inclination angles, $i$, close to the line-of-sight \citep{ree66,zen87,kel04}. Recent claims have shown, however, that a simple ballistic model may explain the observations better if these jetted sources are much closer than their redshifts imply. In the DIR model, quasars are compact, seed objects with a high intrinsic redshift component, that are ejected at all epochs out of the nuclei of mature active galaxies (radio galaxies and Seyferts). As their intrinsic redshift component decreases these compact objects evolve into mature active galaxies in a period of a few times $10^{8}$ yrs \citep{bel02a,bel02b,bel02c,bel02d,bel03,bcr04,bel04,bel06,bel07,mcd06,mcd07}. This model thus differs from the standard model mainly in the way at least some of the galaxies are born and evolve during their first 10$^{8}$ yrs. It is similar in many respects to the local model proposed previously by others from evidence that high-redshift quasars might be associated with low-redshift galaxies \citep{arp97,arp98,arp99,arp01,bur99,gal05,lop06,nar80,nar93}.

In radio-loud jetted sources, because of projection effects in the simple ballistic ejection model, if the ejection speeds are similar, the angular motion $\mu$ (mas yr$^{-1}$) of the ejected blobs is related to the inclination angle, $i$, of the jet relative to the line-of-sight. The largest angular motion is then expected in jets that are in the plane of the sky ($i = 90\arcdeg$). In $\mu$ versus S plots, where S is the radio flux density of the central engine, this scenario then naturally produces an upper cutoff in $\mu$. If the radio luminosity of the central engine is a relatively good standard candle, this upper cut-off will also have a slope of 0.5.



\section{The Data}

In this study we use the source sample discussed by \citet{kel98,kel04} which looked only at the core and inner jet of radio-loud AGN galaxies. This sample offers several advantages that are desirable in this type of examination. The sources were all observed for long periods (many for up to 10 years), using the same observing technique and instrumentation. Furthermore there is less chance that the changes in jet direction or dispersion and possible slowing down in ejection speeds with time seen in kiloparsec jets could have affected these results. In Fig 1 the maximum angular motion seen in the jets of the 114 sources with the most accurate data from this sample is plotted versus the 15 GHz flux density of the central engine. Here the flux density is the maximum value obtained during the observing period. Although the sources were chosen because they have flat spectra, S$_{k}$ represents the measured flux density adjusted by a k-correction factor of (1+z) to correct for bandwidth compression. This correction is necessary regardless of whether the redshift is cosmological or intrinsic. The angular motion in Fig 1 has been normalized to zero redshift using the factor (1+z)$^{2}$ to correct for the change observed in $\mu$ with redshift found previously \citep{mcd07}(see below for further discussion). In Fig 1 an upper limit is detected which can be interpreted in the simple ballistic ejection model as evidence for the maximum velocity expected when ejection is perpendicular to the line-of-sight \citep{bel06,mcd07}. An upper cut-off slope of 0.5 is also observed which can be explained in the DIR model if S$_{15}$ is a good radio standard candle \citep{bel06,mcd07}. As can be seen, the angular motion data cover a range of $\mu$-values from 0.05 to 1 which is compatible with a change in inclination angle from a few degrees to $90\arcdeg$, further supporting the simple ballistic model. In this model the sources near the upper cut-off would be expected to show extended structure while those near the bottom of Fig 1 should show little, or no, extended structure.

\begin{figure}
\hspace{-1.6cm}
\vspace{-1.5cm}
\epsscale{1.1}
\plotone{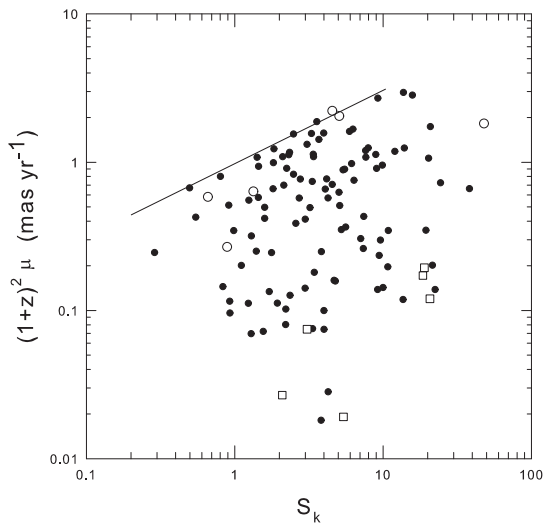}
\caption{{Plot of angular motions in jets versus 15 GHz flux density for sources from \citet{kel04}. The sharp upper cut-off is produced in the DIR model when the jets are in the plane of the sky. The solid line has a slope of 0.5, as is expected if the radio luminosity is a good standard candle. See text for an explanation of the sources plotted as open circles and open squares.
\label{fig1}}}
\end{figure} 
 

The six sources with the largest visible structure in the sample, and with high $\beta_{app}$ values, are shown in Fig 2 and their parameters are listed in Table 1. The six sources with minimum visible structure and low $\beta_{app}$ values are shown in Fig 3 and their parameters are listed in Table 2. The composite maps have been produced from plots taken from \citet{kel98}. Figs 2 and 3 appear to represent sources with large and small viewing angles respectively. Since, in the CR model, the mean distance to the sources in Fig 3 is only a factor of $\sim3$ larger than for the sources in Fig 2, the lack of extended structure in Fig 3 is unlikely to be due solely to a difference in distance. In the simple ballistic ejection scenario proposed in the DIR model, the sources in Fig 2 should then all lie in the top portion of Fig 1, where viewing angles are large, while those in Fig 3 should all be located near the bottom. These two groups of 6 sources have been indicated in Fig 1, where it can be seen that this is the case, with the extended sources (Fig 2) plotted as open circles lying near the top of the plot, and the sources with little extended structure (Fig 3) plotted as open squares lying near the bottom. This result, along with the upper cut-off and slope in Fig 1, \em then can all be easily explained by the simple ballistic ejections of the DIR model, without relativistic motion, Doppler boosting, or superluminal motion. \em

In the CR model highly superluminal sources ($\beta_{app} > 10$) must have their jets directed within $\sim10\arcdeg$ of the line of sight if their superluminal motion is to be explained. Highly superluminal sources might then be expected to show little extended jet structure. In the CR model there is a contradiction in the data. The sources in Fig 2 have high $\beta_{app}$ values $and$ extended structure. Because of their extended structure we expect them to have large viewing angles. But their jets cannot have large viewing angles and be coming towards us at the same time. Thus Fig 1 agrees with what is predicted in the DIR model but presents a contradiction that, at least at first glance, is more difficult to explain in the CR model.


\begin{figure}
\hspace{-1.0cm}
\vspace{-0.0cm}
\epsscale{0.8}
\plotone{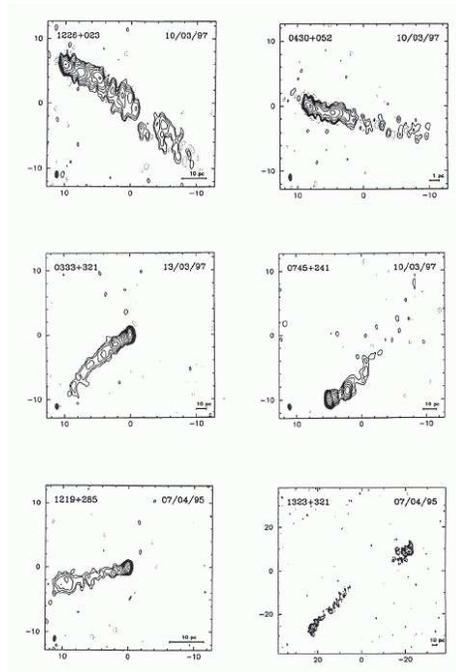}
\caption{{Sources with $\beta_{app} > 3$ and large extended structure selected from \citet{kel98}. \label{fig2}}}
\end{figure}

\begin{figure}
\hspace{-1.0cm}
\vspace{-0.3cm}
\epsscale{0.7}
\plotone{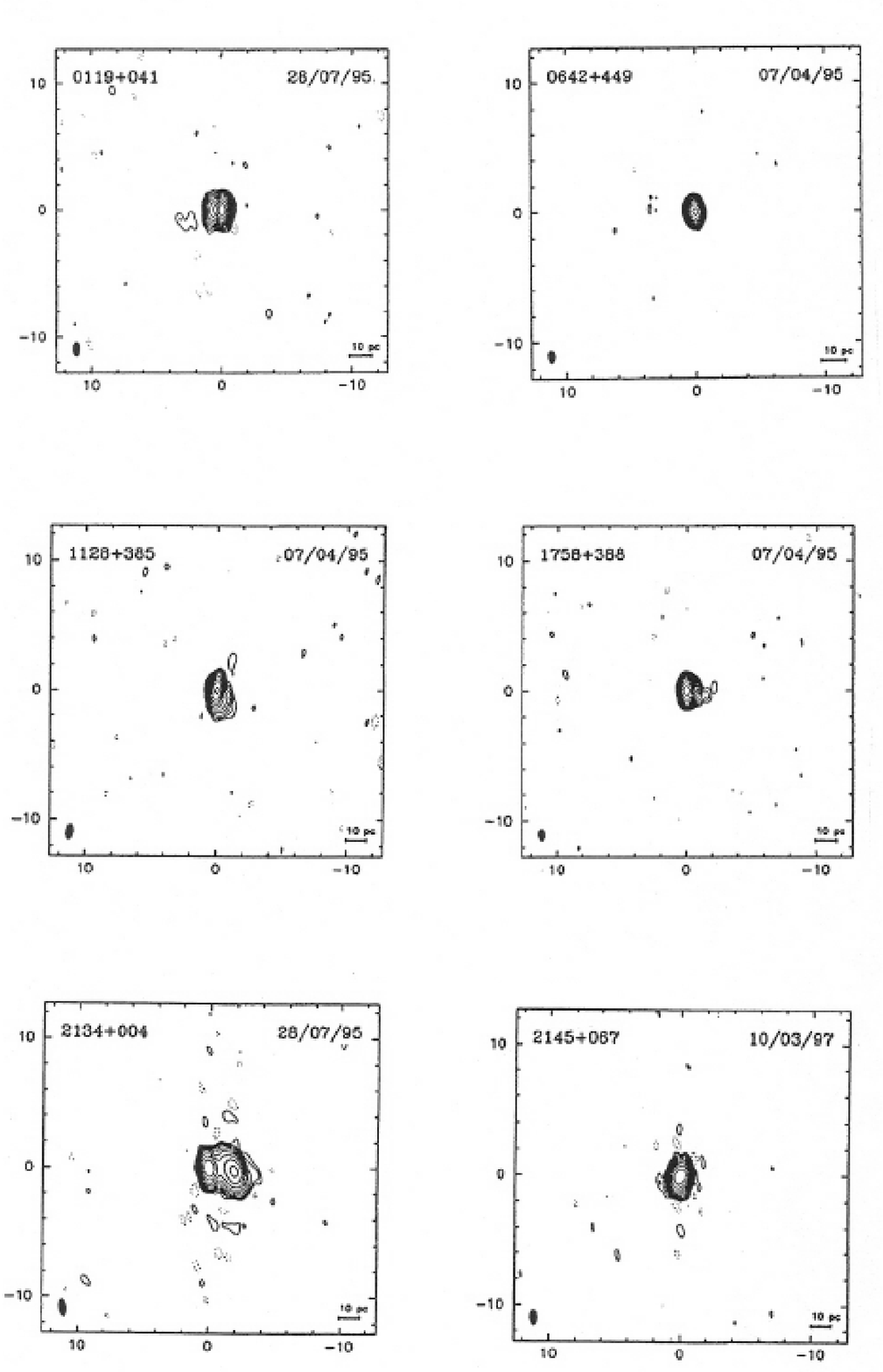}
\caption{{Sources with $\beta_{app} < 1.5$ and little apparent structure selected from \citet{kel98}. \label{fig3}}}
\end{figure}

\section{Discussion}

The relativistic beaming model has been investigated extensively by other authors and will not be discussed in detail here since the purpose of this investigation is see how well the data fit the local model. Possibly because of the large number of parameters it contains ($\mu, \beta, \beta_{p}, \beta_{b}, \beta_{app}, i, \gamma, \gamma_{b}, S, \delta, \tau, v, c, z$), compared to those ($i, \mu, v, c, z, z_{i}$) in the DIR model, the CR has proven to be easily fitted to the data in most cases. But agreement in one model does not rule out another if both can explain the data. 



Since $\beta_{app}$ is directly proportional to $\mu$, when $\beta_{app}$ is plotted vs S$_{k}$ the plot is almost identical to Fig 1. In fact, all the sources with $\beta_{app} > 10$ are located above $\mu(1+z)^{2}$ = 0.75 in Fig 1, as can be seen in Fig 4 where Fig 1 has been replotted with sources with $\beta_{app} > 10$ plotted as open circles. The low-$\beta_{app}$ sources are located near the bottom of the plot. In the DIR model this is acceptable because the apparent superluminal motions are not real, created only by the presence of a large intrinsic component in the redshifts. However, in the CR model, all the high-$\mu$ sources must be coming towards us within $\sim10\arcdeg$ of the line-of-sight if their superluminal motion is to be explained. If the high-$\beta_{app}$ sources near the top of Fig 4 are coming towards us there can be no meaningful relation between $\mu$ and inclination angle in the CR model. Whether in Fig 1 the sharp upper cut-off and its slope of 0.5 can be explained in the CR model is not clear if the jets are coming towards us within 10$\arcdeg$ of the line-of-sight, where the boosting factor is largest and changes quickly with angle. If they cannot, this could be a significant problem for the CR model since these are empirical results that do need to be explained. On the other hand, if they can be explained it would not rule out the DIR model, but it would be an amazing coincidence if all the relations predicted by the simple ballistic model were explainable by the relativistic beaming model. This could then suggest that the many parameters available in the latter model will allow it to be fitted to almost any data.



It is well known that the distribution of angular motions for these sources is not the sin$i$ distribution expected for a random sample of jet orientations. The distribution is highly weighted towards low $\mu$ values \citep[see Fig 22]{bel06}. In the simple ballistic ejection (DIR) model the over-density of sources with small $\mu$ values can be explained without Doppler boosting as follows. The distribution in Fig 1 could result from an anisotropic brightness pattern in which the source radio radiation is \em directed \em in some manner such that the radio brightness increases towards $i = 0\arcdeg$ (small $\mu$) where the core is seen through the hole in the central torus \citep{bel06,mcd07}, keeping in mind that the edge-on sources may also be seen through the host galaxy viewed edge-on. Furthermore, the radio opacity conditions in the regions near the accretion disc are not well known.  From Fig 1 it can be seen that the flux density enhancement at low-$\mu$ values would only need to be a factor of 10-20 to raise all of the low-$\mu$ sources above the $\sim1$ Jy detection limit. The distribution may also be due to the fact that this is not a complete sample, since it was hand picked for core dominant, flat spectrum sources. However, perhaps the most likely reason for the observed $\mu$ distribution is that the distribution is not the result of enhanced radio luminosities at all, but simply due to the fact that most jet motions are at less than the maximum speed. In this case many of the sources with small $\mu$ values can still have large viewing angles. Except for this last case, these explanations are not viable in the CR model because, as shown above, it is the high-$\mu$ sources that must be viewed close to face-on to explain the highly superluminal speeds. (See section 3.2 below for further discussion of this in the CR model.)

In Fig 1 the factor (1+z)$^{2}$ has been used to normalize the $\mu$ values to $z$ = 0. This normalization adjusts for the decrease in $\mu$ with increasing intrinsic redshift reported previously \citep{mcd07}, and which is also shown here in Fig 5. This decrease is now assumed in the DIR model to be due to a significantly lower ejection speed in the younger, high intrinsic redshift objects. Without adjustment for this decrease, objects with large intrinsic redshifts cannot be compared meaningfully to those with small intrinsic redshifts. This normalization then allows us to include the radio galaxies in the same plot with quasars and BL Lacs, while retaining the sharp upper cut-off. Applying this factor of (1+z)$^{2}$ cannot affect the conclusions drawn here from Fig 1 since the high redshift sources (which are affected the most) appear in both source groups. If anything, it is the low-$\mu$ sources (Table 2) that will have moved up the most when this correction was applied, because their mean redshift is slightly higher.
 
An attempt has also been made to explain the decrease and upper cut-off in the $\mu$ versus z plot in the CR model \citep[see their Fig 11]{kel04}. These authors demonstrate how $\mu$ is expected to change with z for a given $\gamma$. However, the fit is crude, as they admit. To obtain a good fit, assuming that a sharp upper cut-off in $\gamma$ can even be expected, this maximum $\gamma$ would have to change with z. There is no obvious reason why there would be a maximum $\gamma$ in the CR model, while in the DIR model the upper cut-off is naturally explained by ejections perpendicular to the line-of-sight if there is a maximum ejection velocity.



\begin{figure}
\hspace{-1.0cm}
\vspace{-1.0cm}
\epsscale{0.9}
\plotone{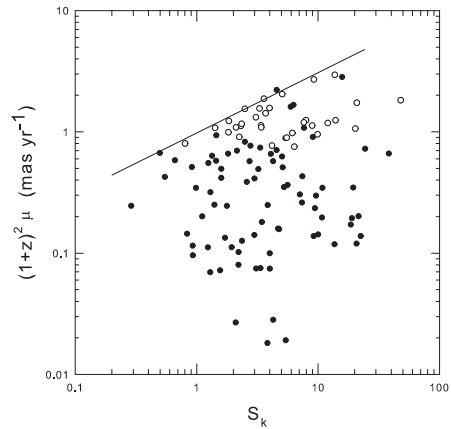}
\caption{{Plot of angular motions in jets versus 15 GHz flux density for sources from \citet{kel04}. Sources plotted as open circles have $\beta_{app} > 10$. 
\label{fig4}}}
\end{figure}

\begin{figure}
\hspace{-1.0cm}
\vspace{-1.0cm}
\epsscale{0.9}
\plotone{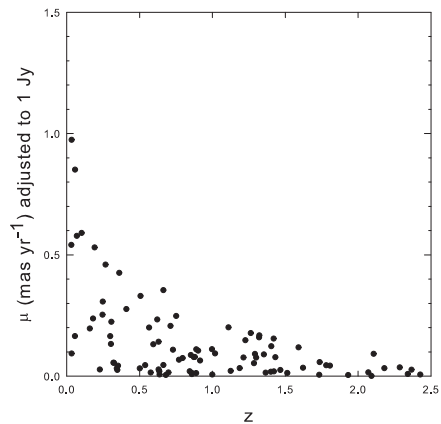}
\caption{{Plot of angular motions in jets normalized to 1 Jy (k-corrected) versus redshift for sources from \citet{kel04}.  
\label{fig5}}}
\end{figure}

    
\begin{figure}
\hspace{-1.0cm}
\vspace{-1.0cm}
\epsscale{1.0}
\plotone{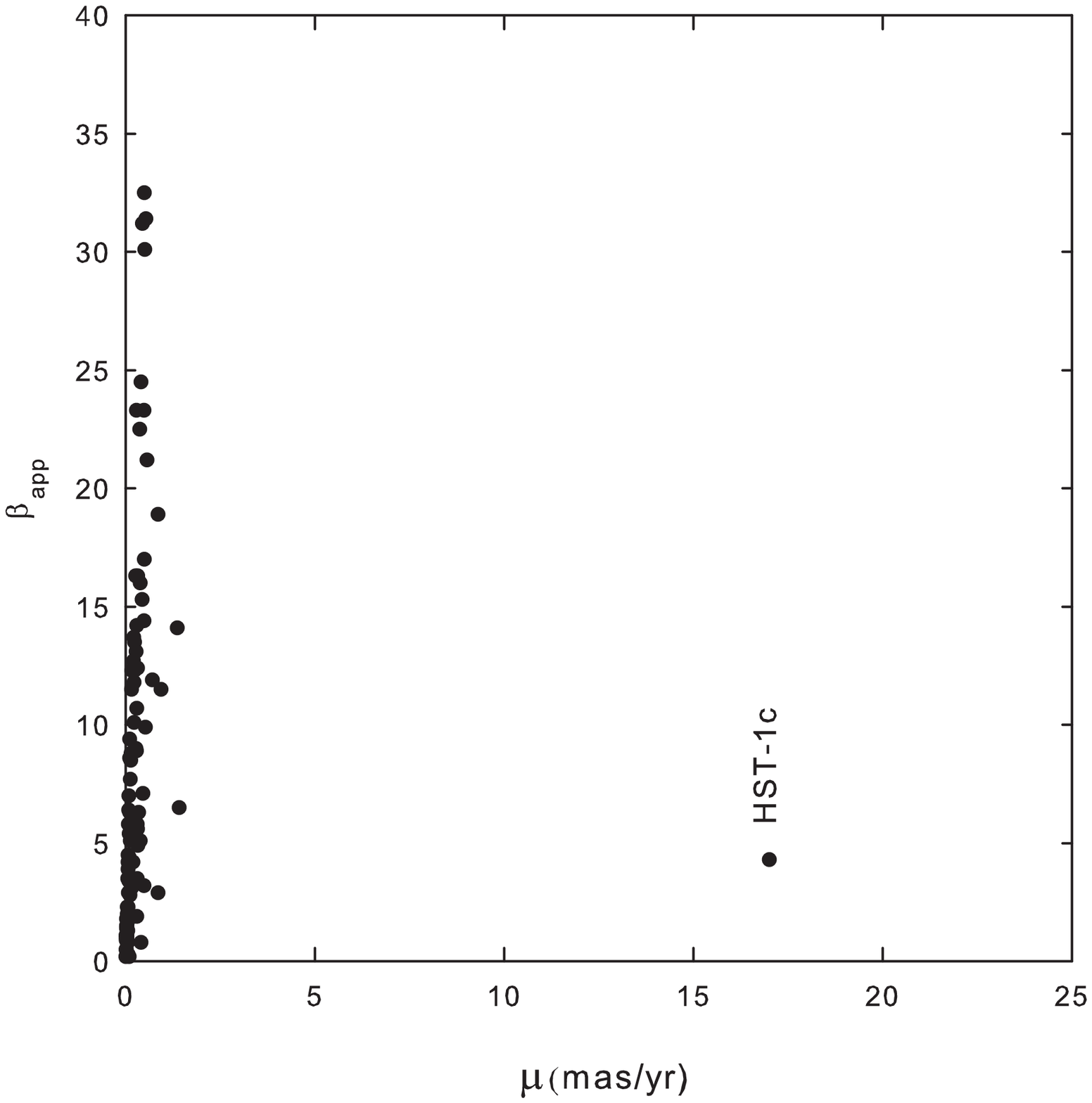}
\caption{{Plot of $\beta_{app}$ versus $\mu$ for sources from \citet{kel04}. The value for HST-1c is from \citet{che07}. 
\label{fig6}}}
\end{figure}

\subsection{Relativistic Ejections in Low-redshift Radio Galaxies}

In the DIR model it is assumed that since $\beta_{app}$ is proportional to distance in the CR model the high $\beta_{app}$ values have been obtained simply because the high redshift sources contain a large intrinsic redshift component. At first glance this would seem to imply that superluminal motion will not be seen in low redshift sources, because they have little intrinsic redshift. This is in fact observed to be the case for the 15 GHz observations being discussed here. However, it is worth pointing out that although no highly superluminal motion is expected in the low-redshift sources, mildly relativistic ejection velocities in radio galaxies are not ruled out in the DIR model. Why this is so is also demonstrated in Fig 5. It is seen that the maximum angular motions (ejection speeds in the DIR model) increase almost exponentially as the intrinsic redshift decreases. The DIR model claims only that there can be no highly relativistic ejection speeds in the jets of high redshift quasars because their maximum ejection speeds are more than an order of magnitude slower than the maximum in mature radio galaxies. The two sources with largest angular motions in Fig 5 are the radio galaxies 3C120 and 3C390.3, which in the CR model have superluminal ejections and in the DIR model could have ejection velocities in excess of 0.5c. In Fig 5, $\mu$ has been normalized to S$_{k}$ = 1 Jy to remove the cut-off smearing effects that would otherwise be introduced by the slope of 0.5 in Fig 1. Note that those sources that define the upper limit in Fig 5 are the same ones that define the upper limit in Fig 1.

It is important to note that the ejection velocities measured by \citet{kel04} are in material ejected from the vicinity of the accretion disc. Higher apparent ejection speeds near 4c-6c have been reported at the so-called HST-1 location in the M87 jet, both at optical \citep{bir99} and at radio \citep{che07} wavelengths. This region appears to be unique in the M87 jet and it has been clearly demonstrated by \citet{che07} that the superluminal motion at this site originates, not in the central compact object like the material observed by \citet{kel04}, but from a point (HST-1d) that is located 120pc downstream in the main jet. Flaring, associated with the superluminal motion seen at this site may also be associated with high energy TeV emission \citep{che07}. This has lead these authors to suggest that the highly superluminal motions detected in more distant objects might also be occurring further downstream, and not in the blobs ejected from near the accretion disc as previously thought. However, it seems unlikely that all of the sources in the Kellermann sample can be explained in this way. In Fig 6, the ejection velocities ($\beta_{app}$) observed by \citet{kel04} are plotted vs angular motions, $\mu$. Included in the plot is the maximum value obtained in HST-1 by \citet{che07}. It seems clear from this plot that the HST-1 value is related to a completely different phenomenon than that producing the Kellermann values.



\subsection{Evidence Against Relativistic Ejection Speeds in the Jets of High-z Quasars}

The relativistic beaming model was devised to explain apparent superluminal motions in quasar jets over 40 years ago \citep{ree66}. Over that period astronomers have accepted this model without confirming proof that it was the correct explanation. Many observations appear to support this model, and one of these is the prevalence of one-sided jets. However, one-sidedness can be explained in other ways and whether or not this is due to Doppler boosting in an approaching jet can only be confirmed by measuring relativistic motions in a source whose accurate distance is known. One such source is M87 (Virgo A) whose distance has been determined independently using Cepheid variables. In its inner jet regions this source also has a jet/counterjet intensity ratio at $\lambda$ 2-cm that is highly asymmetric implying relativistic ejections. Although a concerted effort has recently been made to measure the ejection speeds of this material at radio wavelengths \citep{kov07}, surprisingly no ejection speeds in excess of a few percent of the speed of light have been found in blobs that originate near the central compact object and show the jet/counterjet asymmetry. This has forced \citet{kov07} to conclude that the asymmetry in intensities may be intrinsic, and not due to Doppler boosting. \em This is what is expected in the DIR model, but it is not expected in the relativistic beaming model. \em

The presence of Doppler boosting, or lack of it, can also be tested using the distribution of angular motions. As noted above, in radio-loud AGN jets the distribution of angular motions is known to be heavily weighted to small $\mu$ values \citep{kel04} instead of the sin$i$ relation expected for a random distribution \citep[see Fig 22]{bel06}. This abnormal distribution was eagerly embraced by astronomers when it was thought that the low $\mu$ values indicated that the jets were coming towards us and the high source density at low $\mu$ values was then naturally explained by Doppler boosting. However, because the high-$\beta_{app}$ sources are associated with sources near the top in Fig 4 it would seem that this can no longer be the case. It is the high-$\mu$ sources that must be coming towards us because they have high $\beta_{app}$ values, and therefore must be Doppler boosted in the CR model, while the low-$\mu$ sources do not show large superluminal motions. What now needs to be explained in the CR model is why the high $\mu$, and supposedly highly Doppler boosted, sources are so few in number while those with little superluminal motion, or Doppler boosting, are so prevalent. A similar result is obtained directly from the $\mu$-distribution without converting to inclination angle. This suggests, as above, that Doppler boosting does not play a role.

The superluminal velocities seen in HST-1 do not alter the results obtained by \citet{kov07}.


 \section{Conclusions}

An examination of the structure and angular motions in the inner jets of radio loud quasars and other AGN galaxies reveals that the simple ballistic ejections proposed in the DIR model can easily explain the data, showing good agreement with what is predicted by the model. Not only are the jets with extended structure likely to be close to the plane of the sky, and those with little structure likely to be coming towards us, an upper limit with a slope of 0.5 is also seen in the $\mu$ vs S plot, as is predicted in this model if the radio luminosity of the central engine is a good standard candle. Whether the abrupt upper cut-off and slope of 0.5 can be explained in the relativistic beaming model is at present unclear. In the CR model the distribution of radio-loud AGN sources is also found to be heavily weighted towards sources that are not significantly boosted while those that are highly Doppler boosted are scarce. This is opposite to what is expected and suggests that Doppler boosting may not play a role. When this is taken together with the recent results of \citet{kov07}, it again implies that the DIR model may explain the data best.

\clearpage

\begin{deluxetable}{ccccccc}
\tabletypesize{\scriptsize}
\tablecaption{Sources With Extended Structure and $\beta_{app} > 3$. \label{tbl-1}}
\tablewidth{0pt}
\tablehead{
\colhead{Source} & \colhead{Type} & \colhead{$\mu$(mas yr$^{-1}$)} & \colhead{$\mu$(1+z)$^{2}$} & \colhead{Redshift} & \colhead{S$_{k}$} & \colhead{$\beta_{app}$}
}

\startdata

0333+321 & Q & 0.40$\pm0.07$ & 2.048 & 1.263 & 5.07 & 24.5\\
0430+052(3C120) & G & 2.08$\pm0.24$ & 2.22 & 0.033 & 4.55 & 4.6 \\
0745+241 & Q & 0.32$\pm0.05$ & 0.64 & 0.409 & 1.34 & 7.9 \\
1219+285 & B & 0.48$\pm0.03$ & 0.582 & 0.102 & 0.66 & 3.2 \\
1226+023 & Q & 1.36$\pm0.11$ & 1.82 & 0.158 & 47.94 & 14.1 \\
1323+321 & G & 0.142$\pm0.001$ & 0.266 & 0.37 & 0.89 & 3.25 \\

\enddata 

\end{deluxetable}

\begin{deluxetable}{ccccccc}
\tabletypesize{\scriptsize}
\tablecaption{Sources With No Extended Structure and $\beta_{app} < 1.5$. \label{tbl-2}}
\tablewidth{0pt}
\tablehead{
\colhead{Source} & \colhead{Type} & \colhead{$\mu$(mas yr$^{-1}$)} & \colhead{$\mu$(1+z)$^{2}$} & \colhead{Redshift}  & \colhead{S$_{k}$} & \colhead{$\beta_{app}$}
}

\startdata

0119+041 & Q & 0.01 & 0.027 & 0.637 & 2.09 & 0.5 \\
0642+449 & Q & 0.01 & 0.194 & 3.408 & 19.0 & 0.9 \\
1128+385 & Q & 0.01 & 0.075 & 1.733 & 3.08 & 1.1 \\
1758+358 & Q & 0.002 & 0.019 & 2.092 & 5.41 & 0.2 \\
2134+004 & Q & 0.02 & 0.172 & 1.932 & 18.59 & 1.5 \\
2145+067 & Q & 0.03 & 0.120 & 0.999 & 20.72 & 1.4 \\

\enddata 

\end{deluxetable} 

\clearpage


\clearpage
\begin{thebibliography}

\bibitem[Arp(1997)]{arp97} Arp, H. 1997, \aap, 319, 33
\bibitem[Arp(1998)]{arp98} Arp, H. 1998, \apj, 496, 661 
\bibitem[Arp(1999)]{arp99} Arp, H. 1999, \apj, 525, 596
\bibitem[Arp and Russell(2001)]{arp01} Arp, H., and Russell, D.G. 2001, \apj, 549, 802
\bibitem[Bell(2002a)]{bel02a} Bell, M.B. 2002a, \apj, 566, 705
\bibitem[Bell(2002b)]{bel02b} Bell, M.B. 2002b, \apj, 567, 801
\bibitem[Bell(2002c)]{bel02c} Bell, M.B. 2002c, astro-ph/0208320
\bibitem[Bell(2002d)]{bel02d} Bell, M.B. 2002d, astro-ph/0211091
\bibitem[Bell(2004)]{bel04} Bell, M.B. 2004, \apj, 616, 738
\bibitem[Bell and Comeau(2003)]{bel03} Bell, M.B. and Comeau, S.P. 2003, astro-ph/0305060
\bibitem[Bell, Comeau and Russell(2004)]{bcr04} Bell, M.B., Comeau, S.P. and Russell, D.G. 2004, astro-ph/0407591
\bibitem[Bell(2006)]{bel06} Bell, M. B. 2006, astro-ph/0602242
\bibitem[Bell(2007)]{bel07} Bell, M. B. 2007, \apj, 667, L129 (arXiv:0704.1631)
\bibitem[Bell and McDiarmid(2006)]{mcd06} Bell, M.B. and McDiarmid, D. 2006, ApJ, 648,140, and astro-ph/0603169
\bibitem[Bell and McDiarmid(2007)]{mcd07} Bell, M.B. and McDiarmid, D. 2007, astro-ph/0701093
\bibitem[Biretta et al.(1999)]{bir99} Biretta, J.A. Sparks, W.B., and Macchetto, F 1999, \apj, 520, 621
\bibitem[Burbidge(1999)]{bur99} Burbidge, E.M. 1999, \apj, 511, L9
\bibitem[Cheung et al.(2007)]{che07} Cheung, C.C., Harris, D.E. and Stawarz, L. 2007, \apj, 663, L65
\bibitem[Galianni et al.(2005)]{gal05} Galianni, P. et al. 2005, \apj, 670, 88
\bibitem[Kellermann et al.(1998)]{kel98} Kellermann, K.I., Vermeulen, R.C., Zenus, J.A., and Cohen, M.H. 1998, \aj, 115, 1295
\bibitem[Kellermann et al.(2004)]{kel04} Kellermann, K.I. et al. 2004, \apj, 609, 539
\bibitem[Kovalev et al.(2007)]{kov07} Kovalev, Y.Y., Lister, M.L. Homan, D.C., and Kellermann, K.I. 2007, \apj, 668, L27
\bibitem[Lop$\acute{e}$z-Corredoira and Guti$\acute{e}$rrez (2006)]{lop06} Lop$\acute{e}$z-Corredoira, M. and Guti$\acute{e}$rrez, C.M. 2006, astro-ph0609514
\bibitem[Narlikar and Das(1980)]{nar80} Narlikar, J.V. and Das, P.K. 1980, \apj, 240,401
\bibitem[Narlikar and Arp(1993)]{nar93} Narlikar, J.V. and Arp, H. 1993, \apj, 405, 52
\bibitem[Rees(1966)]{ree66} Rees, M.J. 1966, Nature, 211, 468
\bibitem[Zensus and Pearson(1987)]{zen87} Zensus, J.A. and Pearson, T.J. 1987, in \em Superluminal Radio Sources, \em  eds. Zensus and Pearson, 1987, Cambridge University Press

\end{thebibliography}
\end{document}